\documentclass[
 aip,
 amsmath,amssymb,
 reprint
]{revtex4-1}

\usepackage{graphicx}
\usepackage{dcolumn}
\usepackage{bm}
\usepackage[utf8]{inputenc}
\usepackage[T1]{fontenc}
\usepackage{mathptmx}
\usepackage{etoolbox}

\makeatletter
\def\@email#1#2{%
 \endgroup
 \patchcmd{\titleblock@produce}
  {\frontmatter@RRAPformat}
  {\frontmatter@RRAPformat{\produce@RRAP{*#1\href{mailto:#2}{#2}}}\frontmatter@RRAPformat}
  {}{}
}%
\makeatother

\begin{document}

\preprint{AIP/123-QED}

\title{Energy--time entanglement from a monolithically integrated quantum dot on silicon}

\author{Marcel Hohn}
\thanks{These authors contributed equally to this work.}
\affiliation{Institute for Physics and Astronomy, Technical University of Berlin, Hardenbergstraße 36, D-10623 Berlin, Germany}
\thanks{These authors contributed equally to this work.}

\author{Imad Limame*}
\thanks{These authors contributed equally to this work.}
\email{imad.limame@tu-berlin.de}
\affiliation{Institute for Physics and Astronomy, Technical University of Berlin, Hardenbergstraße 36, D-10623 Berlin, Germany}

\author{Peter Ludewig}
\affiliation{mar.quest | Marburg Center for Quantum Materials and Sustainable Technologies, Philipps University Marburg, 35032 Marburg, Germany}
\affiliation{Department of Physics, Philipps University Marburg, Hans Meerwein Str. 6, 35032 Marburg, Germany}

\author{Chirag C. Palekar}
\affiliation{Institute for Physics and Astronomy, Technical University of Berlin, Hardenbergstraße 36, D-10623 Berlin, Germany}

\author{Aris Koulas-Simos}
\affiliation{Institute for Physics and Astronomy, Technical University of Berlin, Hardenbergstraße 36, D-10623 Berlin, Germany}

\author{Kerstin Volz}
\affiliation{mar.quest | Marburg Center for Quantum Materials and Sustainable Technologies, Philipps University Marburg, 35032 Marburg, Germany}
\affiliation{Department of Physics, Philipps University Marburg, Hans Meerwein Str. 6, 35032 Marburg, Germany}

\author{Stephan Reitzenstein}
\affiliation{Institute for Physics and Astronomy, Technical University of Berlin, Hardenbergstraße 36, D-10623 Berlin, Germany}

\date{\today}

\begin{abstract}
Scalable quantum photonic technologies require deterministic sources of entangled photons that are compatible with established semiconductor manufacturing platforms. While self-assembled III--V semiconductor quantum dots are among the most promising sources of on-demand entanglement generation, their integration with silicon-based architectures remains a central challenge. Here, we demonstrate energy--time entanglement from a single InGaAs/GaAs quantum dot monolithically grown on a silicon substrate. Under coherent two-photon excitation, we achieve coherent control of the biexciton--exciton cascade, evidenced by Rabi oscillations and dressed-state formation. Using a four-channel Franson interferometer, we observe phase-dependent two-photon interference with visibilities up to $(64.0 \pm 7.0)\%$ for an 80 ps integration window (and $(49.4 \pm 1.9)\%$ for a 1600 ps window), approaching the threshold for Bell inequality violation at short time scales. These results establish monolithically integrated III--V-on-silicon quantum dots as promising sources of energy--time entangled photons for scalable quantum photonic technologies.
\end{abstract}

\maketitle

\section{Introduction}

Semiconductor III--V quantum dots (QDs) have emerged as a leading solid-state platform for quantum photonic technologies based on the generation of indistinguishable and entangled single photons~\cite{Senellart2017, Heindel2023, Esmann2024, Xavier2025}. Over the past decade, QDs have demonstrated near-ideal quantum optical performance, including high single-photon purity~\cite{Schweickert2018}, high entanglement fidelity~\cite{Huber2016,Chen2024}, and high photon indistinguishability~\cite{Reitzenstein_CSSCST2024}, while simultaneously achieving high extraction efficiencies when integrated into photonic nanostructures~\cite{Ding2016,Yang:24}. Importantly, their emission wavelength can be engineered from the near-infrared to the telecom C-band~\cite{Sartison2018,Hauser2026}, making them highly versatile for both fiber-based and on-chip quantum networks.

These exceptional optical properties have positioned QDs as key quantum light sources for a range of quantum information protocols, including quantum key distribution (QKD)~\cite{Waks2002, Heindel2012}, boson sampling~\cite{Loredo2017,Wang2018}, and entanglement swapping~\cite{BassoBasset2019,Zopf2019}. In parallel, substantial progress has been made toward scalable device architectures through deterministic QD positioning techniques. Approaches such as site-controlled growth using buried stressors~\cite{Strittmatter2012,Gaur2026}, nanohole-template-assisted nucleation~\cite{Schneider2012}, inverted pyramids~\cite{PELUCCHI2004476}, and post-growth deterministic integration via in-situ or marker-based lithography~\cite{Dousse2008,Gschrey2013,He2017} have enabled precise spatial control of emitters, even facilitating the realization of integrated quantum photonic circuits \cite{Li2023}. Furthermore, the monolithic integration of III--V photonic devices---including lasers, detectors, and, as demonstrated here, quantum emitters---on silicon represents a major milestone toward the seamless integration of silicon photonics with III--V nanophotonic platforms for both classical and quantum technologies \cite{Limame2024, Koscica2025}. In the context of InGaAs/GaAs QDs, the introduction of a GaP intermediate buffer layer on silicon substrate has enabled the epitaxial growth of high-quality single-photon-emitting QDs operating both in the near-infrared around 930~nm and in the telecom O-band~\cite{Limame2024,Tripathi2026}.

A central feature of semiconductor QDs is their ability to generate entangled photon pairs via the biexciton-exciton (XX--X) radiative cascade \cite{Benson2000}. To date, most investigations of entangled-photon generation in QDs have focused on polarization entanglement \cite{Akopian2006,Young_2006, Huber2017,Chen2024}. However, the achievable entanglement fidelity is fundamentally limited by the fine-structure splitting (FSS) of the exciton state, which originates from structural asymmetries and inhomogeneous strain distributions introduced during the growth process. Although significant progress has been made toward reducing the FSS through approaches such as droplet-etched GaAs/AlGaAs QDs \cite{Huo2013} and the growth of InGaAs QDs on (111)-oriented GaAs substrates \cite{Yerino2014, vonHelversen2022}, these strategies often involve trade-offs, including restricted emission wavelength tunability or reduced optical quality.

Energy-time entanglement provides an attractive alternative that is intrinsically insensitive to FSS. In this approach, entanglement is encoded in the temporal and spectral correlations of photon pairs rather than in their polarization degree of freedom, thereby circumventing symmetry-related limitations \cite{Sun2017,Hohn2023}. While energy-time entanglement has been extensively studied using spontaneous parametric down-conversion (SPDC) sources \cite{Xavier2025}, the probabilistic nature of SPDC fundamentally limits source brightness, scalability, and, in some quantum communication protocols, security. Semiconductor QDs, by contrast, offer a deterministic route to the generation of entangled photon pairs. Energy--time entanglement from QDs has been experimentally demonstrated using time-bin and Franson-type interferometric measurements, including full quantum state tomography \cite{Jayakumar2014,Gins2021}. Furthermore, the generation of multi-photon time-bin-entangled states has been reported \cite{Lee2019}, highlighting the potential of QDs as scalable sources for advanced quantum photonic protocols.

Here, we report on the generation of energy-time entangled photon pairs under coherent excitation from InGaAs QDs integrated monolithically on silicon substrate. Energy–time interference measurements under two-photon excitation (TPE) yield Franson visibilities of up to $(64 \pm 7)\%$ for an 80 ps integration window, demonstrating non-classical energy–time correlations while remaining below the threshold required for a Bell-inequality violation \cite{Bell1964}. The remaining limitations in visibility are attributed to residual non-interfering contributions rather than to fundamental constraints of the platform, highlighting silicon-integrated QDs as a promising route toward scalable, silicon-compatible quantum photonic technologies.

\section{Monolithic integration on silicon}

\begin{figure*}[tbp]
\centering
  \includegraphics[height=8cm]{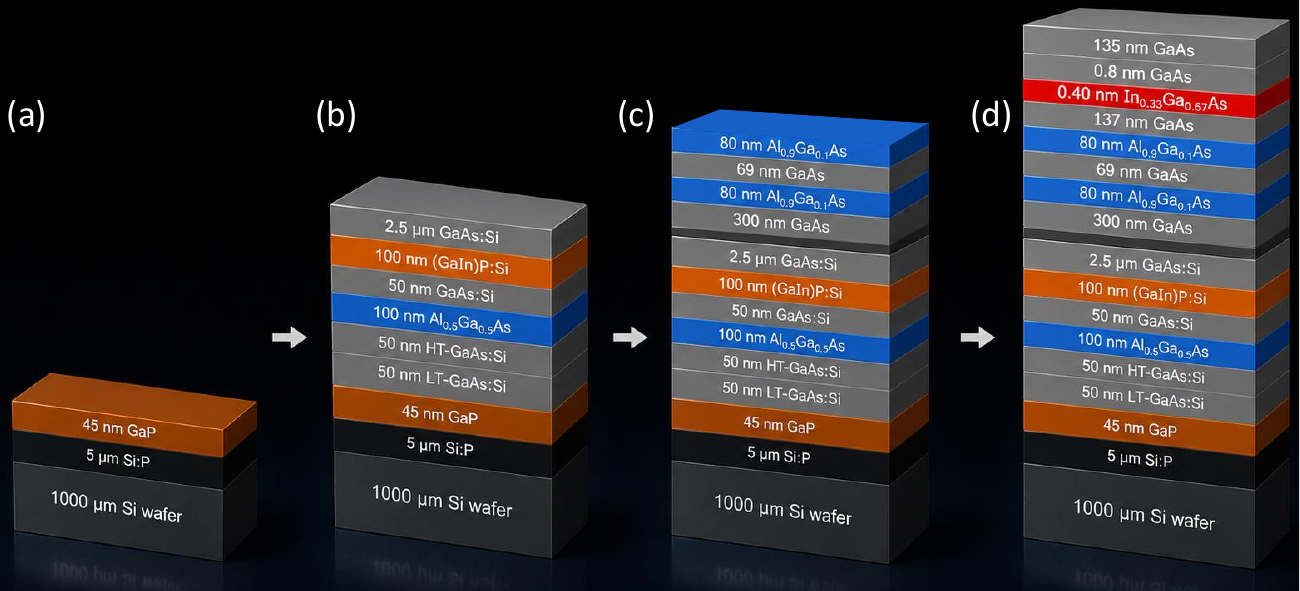}
  \caption{Schematic of the epitaxial layer structure grown on a (001)-oriented Si substrate. The lower stack comprises a Si-compatible GaP nucleation layer (a), followed by GaAs-, AlGaAs-, and GaInP-based layers for strain accommodation and dislocation reduction (b), culminating in a high-quality virtual GaAs substrate. The upper structure consists of a GaAs buffer, a backside DBR (32.5 pairs of GaAs/Al$_{0.90}$Ga$_{0.10}$As) (c), and an active region formed by self-assembled In$_{0.33}$Ga$_{0.67}$As QDs embedded within a $\lambda/n$ GaAs cavity (d).\cite{ChatGPT2026}}
  \label{Fig:Epi}
\end{figure*}

The investigated QD-heterostructure was grown by metal--organic chemical vapor deposition (MOCVD) in a two-step epitaxial process, schematically illustrated in Fig.~\ref{Fig:Epi} and described in detail in Ref.~\cite{Limame2024}. In the first step, an exactly oriented 300~mm Si~(001) substrate was overgrown in an AIXTRON Crius R CCS twin-reactor system with a 5~$\mu$m Si:P buffer layer, followed by the deposition of a GaP III--V nucleation layer \cite{Volz2011}. This nucleation layer serves as a critical interface between the non-polar Si substrate and the polar III--V material system, thereby suppressing the formation of anti-phase domains\cite{Beyer2013}. Subsequently, a sequence of AlGaAs/GaAs and GaInP dislocation-filtering layers was deposited, acting both as lattice-accommodating layers and as an effective means to reduce the threading dislocation density (TDD). These layers facilitate dislocation bending and annihilation, significantly improving the crystalline quality of the III/V heterostructure \cite{Volz2011,Beyer2013}. The first epitaxial stage was completed by the growth of a 2.5~$\mu$m GaAs layer, which provides a high-quality virtual GaAs substrate suitable for subsequent QD epitaxy (Fig.~\ref{Fig:Epi}, bottom). Following this initial growth, the wafer was cleaved into $4 \times 4~\mathrm{cm}^2$ samples and transferred to an AIX 200/4 MOCVD reactor for the second epitaxial step. The second epitaxial stage starts with the deposition of a 300~nm GaAs buffer layer at 700~°C under a high V/III ratio of 200, ensuring excellent crystalline quality. Subsequently, 32.5 periods of Al$_{0.90}$Ga$_{0.10}$As/GaAs are grown at the same temperature, forming a highly reflective bottom distributed Bragg reflector (DBR) designed to enhance photon out-coupling into the 0.69~NA collection optics of the $\mu$PL setup. The individual layer thicknesses correspond to $\lambda/4$, with 68~nm for GaAs and 81~nm for Al$_{0.90}$Ga$_{0.10}$As. The structure is completed by a $\lambda/n$ GaAs cavity (274~nm) incorporating a 0.40~nm In$_{0.33}$Ga$_{0.67}$As seed layer at its center. A growth interruption of 40~s is applied to promote the self-assembly of QDs, yielding an areal density on the order of $1 \times 10^{9}~\mathrm{cm}^{-2}$. We refer to Ref.~\cite{Limame2024} for details on the structural quality and surface morphology of the grown sample.

\begin{figure*}
 \centering
 \includegraphics[height=8cm]{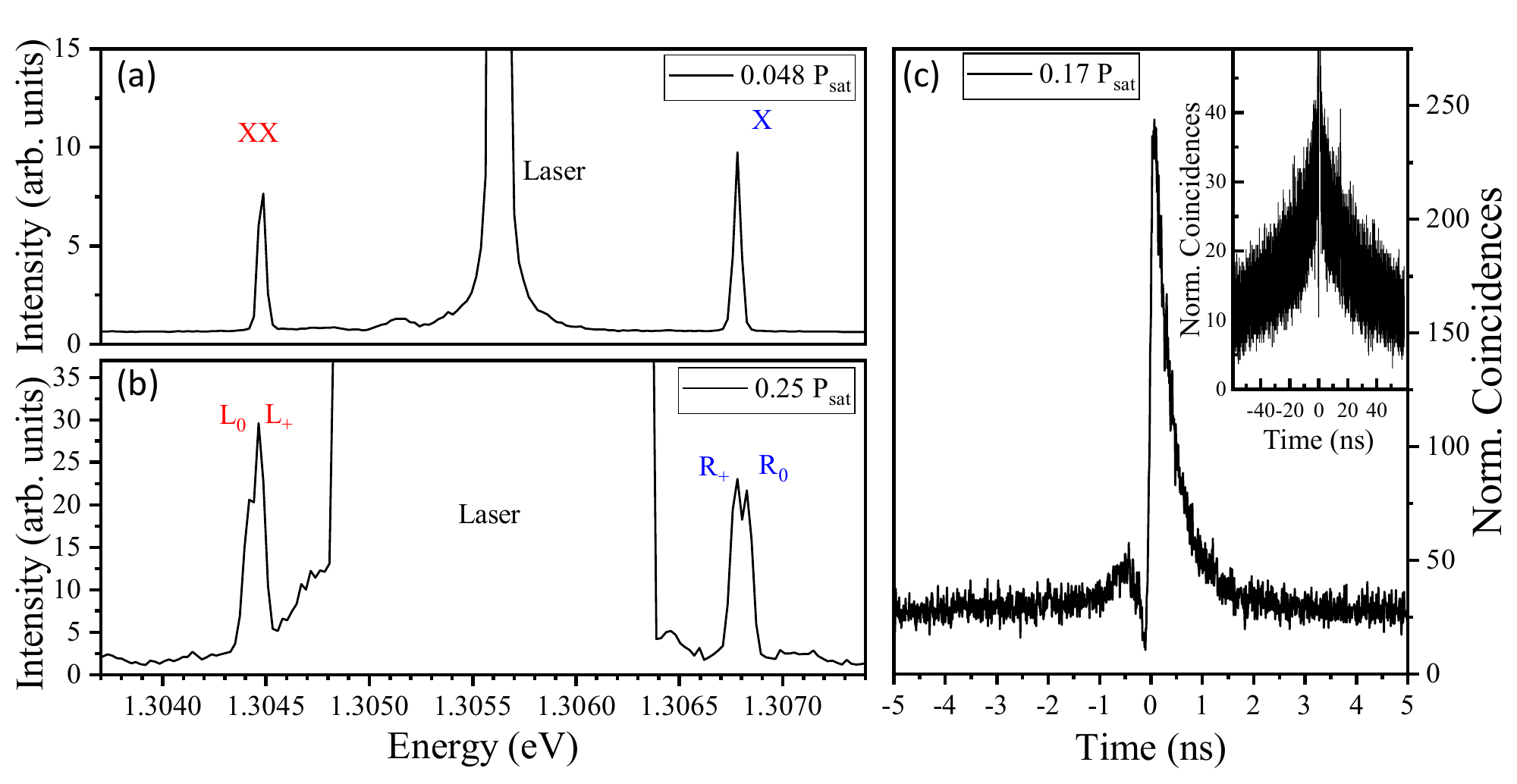}
 \caption{Emission spectra of the investigated QD under CW resonant TPE at excitation powers of $0.048\,P_{\mathrm{sat}}$ (a) and $0.25\,P_{\mathrm{sat}}$ (b). At elevated power (b), the emergence of dressed-state splitting evidences coherent driving of the transition. Laser suppression is achieved via polarization filtering in a cross-polarized configuration. (c) Cross-correlation histogram between biexciton (XX) and exciton (X) emission at $0.17\,P_{\mathrm{sat}}$. The pronounced antibunching at negative delays and bunching at positive delays unambiguously confirm the cascaded emission process. The observed Rabi oscillations further substantiate coherent excitation. Inset: long-timescale bunching attributed to blinking and re-excitation dynamics.}
 \label{Fig:TPE}
\end{figure*}

\section{Coherent biexciton excitation and verification of the biexciton-exciton cascade}

Following epitaxial growth, the sample is mounted on a three-axis piezoelectric stage inside a closed-cycle helium cryostat and cooled to 4~K. Optical excitation is provided by a tunable continuous-wave (CW) laser and the QD emission is collected in a confocal configuration using a 0.69~NA aspheric lens. The emission spectra are recorded with a silicon CCD coupled to a monochromator equipped with a 1500~lines/mm grating. Time-resolved photoluminescence (TRPL) and second-order autocorrelation measurements, $g^{(2)}(\tau)$, are performed by detecting spectrally filtered QD emission with superconducting nanowire single-photon detectors (SNSPDs) providing a temporal resolution of 30 ps. More details on the used $\mu$PL setup are provided in Ref. ~\cite{Limame2024}.

Using power- and polarization-resolved $\mu$PL measurements we identified the exciton (X) and biexciton (XX) transitions, and the investigated QD is excited via TPE, as shown in Fig.~\ref{Fig:TPE}. Panels~(a) and (b) present representative $\mu$PL spectra acquired at excitation powers of $0.048\times P_{\mathrm{sat}}$ and $0.25\times P_{\mathrm{sat}}$, respectively. At low excitation power, the X and XX emission lines are observed at approximately 1.3045~eV (950.43 nm) and 1.3068~eV (948.76 nm). The excitation laser is tuned to 1.3055~eV, such that two laser photons resonantly drive the two-photon transition from the ground state to the biexciton state via a virtual intermediate state. As the excitation power is increased, the resulting XX-X cascade emission becomes increasingly pronounced. Since the excitation wavelength lies outside the suppression range of the available notch filters, The residual laser light is rejected by cross-polarized detection and spatial mode filtering through the single-mode collection fiber. At elevated excitation powers, the emergence of dressed-state splitting (Fig.~\ref{Fig:TPE}(b)) provides clear evidence of coherent driving of the QD \cite{Jundt2008,SnchezMuoz2015}. In particular, the splitting into the $L_{+,0}$ and $R_{+,0}$ states is resolved in the cross-polarized detection configuration. The observed dressed-state splitting provides direct evidence of coherent light--matter interaction and enables extraction of the Rabi frequency, which scales linearly with the square root of the excitation power.

The cross-correlation of the XX--X cascade was investigated at an excitation power of $0.17\times P_{\mathrm{sat}}$, and the resulting histogram of coincidence counts is shown in Fig.~\ref{Fig:TPE}(c). The histogram exhibits pronounced antibunching at negative delay times, corresponding to the arrival of single biexciton (XX) photons, followed by strong bunching at positive delays associated with the subsequent exciton (X) emission. This asymmetric correlation signature unambiguously confirms the cascaded nature of the XX--X radiative decay process. The inset reveals an additional long-time bunching behavior, which is typical for semiconductor QDs and attributed to blinking dynamics and re-excitation processes within the QD system \cite{Davano2014}.

\section{Intrinsic temporal and coherence constraints of the biexciton-exciton cascade}

In this section, we assess the coherence properties and temporal dynamics of the XX--X cascade relevant for energy--time entanglement experiments, TRPL measurements were performed under off-resonant pulsed excitation at 1.3931~eV (890~nm) using a tunable optical parametric oscillator (OPO) delivering 2~ps pulses at a repetition rate of 80~MHz. The extracted decay times are $(574 \pm 45)$~ps for the X transition and $(286 \pm 12)$~ps for the XX transition, consistent with the expected radiative lifetime relation between the two states, $T_{1,\mathrm{X}} \approx 2T_{1,\mathrm{XX}}$. Both lifetimes are substantially shorter than the radiative decay times of approximately 1-1.5~ns typically reported for comparable InGaAs/GaAs QDs grown on GaAs substrates and on silicon without Purcell enhancement\cite{Limame2024}. This comparatively short radiative lifetimes observed in the present sample are likely a consequence of the relatively low nominal indium concentration (33\%), which promotes enhanced electron--hole wavefunction overlap and, consequently, an increased oscillator strength, in agreement with previous reports~\cite{Reithmaier2004}. Franson interference requires an interferometer delay for which the indistinguishable short-short (SS) and long-long (LL) two-photon amplitudes overlap within the selected coincidence window, while the short-long (SL) and long-short (LS) contributions remain temporally distinguishable. In addition, the two-photon coherence must be preserved over the interferometer imbalance. The intrinsic time ordering and lifetime hierarchy of the XX--X cascade therefore impose an important constraint: for a given interferometer delay, the temporal overlap between the SS and LL amplitudes decreases as the biphoton wavepacket becomes short compared to the path imbalance. Consequently, the achievable Franson visibility is highly sensitive to the combined influence of radiative lifetimes, coherence times, temporal postselection window, and interferometer delay. While energy--time entanglement is generally less sensitive to FSS than polarization entanglement, residual exciton splitting can still indirectly affect phase stability in interferometric measurements under finite temporal resolution. Nevertheless, the cascade ordering remains the dominant intrinsic limitation, as it fundamentally restricts the temporal overlap between the SS and LL pathways and thereby sets an upper bound on the achievable Franson interference visibility, which become particularly sensitive to the coherence properties of the cascade.

Additionally, the single-photon coherence properties of the X and XX transitions were investigated using Michelson interferometry under TPE at an excitation power of $0.17\times P_{\mathrm{sat}}$. The extracted coherence times are $T_{2,\mathrm{XX}} = (128 \pm 4)$~ps and $T_{2,\mathrm{X}} = (91 \pm 3)$~ps, respectively. For both transitions, $T_2 < 2T_1$ indicating the presence of pure dephasing in addition to radiative decay. Interestingly, the longer coherence time observed for the XX transition suggests a higher degree of phase preservation, consistent with its direct coherent excitation via TPE, whereas the X state is populated indirectly through spontaneous radiative decay in the cascade. It is further noted that the measured coherence times exhibit a pronounced dependence on excitation power, as reduced excitation conditions mitigate laser-induced decoherence mechanisms such as charge noise activation, local heating, and re-excitation processes.

\section{Franson interferometry}

\begin{figure*}[tbp]
 \centering
 \includegraphics[height=7cm]{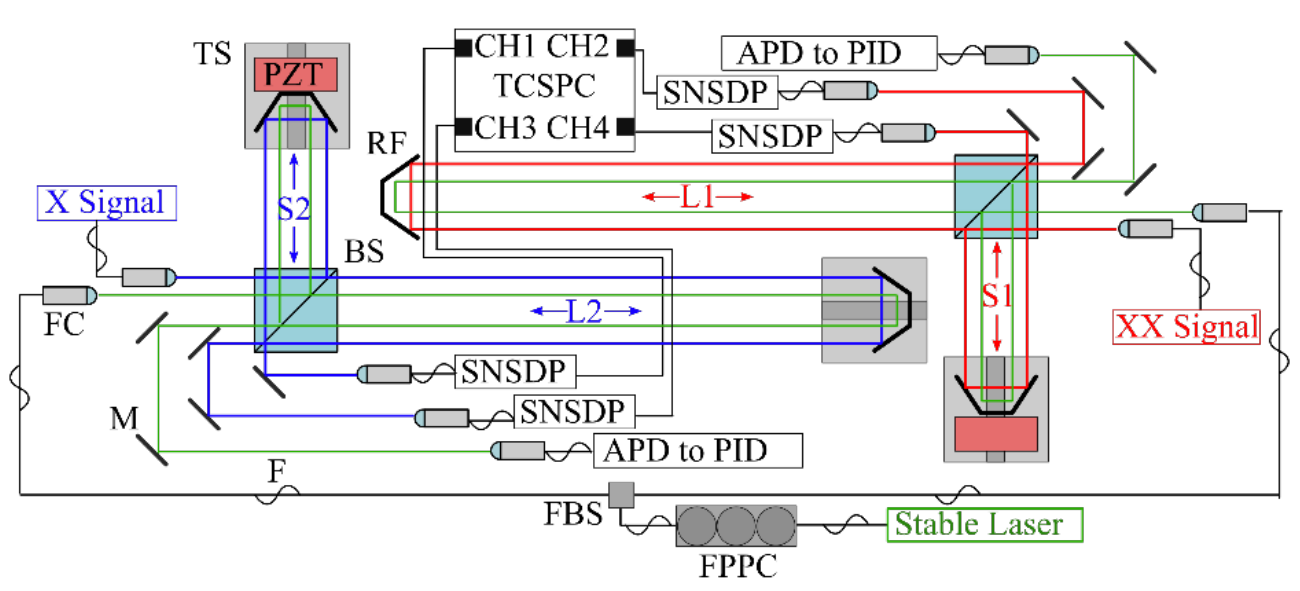}
 \caption{Four-channel Franson interferometer configuration: XX and X photons are spectrally separated by a transmission spectrometer and fiber-coupled into two unbalanced Michelson interferometers (red and blue arms). Each interferometer comprises a non-polarizing 50/50 beamsplitter and a pair of retroreflectors, with the short-arm mirrors mounted on piezoelectric actuators enabling precise phase control. A co-propagating reference laser (green) is employed for active phase stabilization of both interferometers.}
 \label{Fig:Setup}
\end{figure*}

In this section, we first describe the implemented Franson interferometer setup (Fig.~\ref{Fig:Setup}), followed by the corresponding measurements performed on the QD under study. 

The experiment is based on two unbalanced Michelson interferometers, mounted on a vibration-isolated optical table and enclosed in a housing that suppresses stray light and air fluctuations. Although the interferometric setup operates in free space, all input and output ports are fiber-coupled to enable a modular configuration compatible with a transmission spectrometer, which provides spectral separation of the XX (red) and X (blue) emission, as well as fiber-based single-photon detection, as displayed in Fig.~\ref{Fig:Setup}. Each interferometer arm consists of a non-polarizing 50/50 beamsplitter and two retroreflectors. The use of retroreflectors instead of planar mirrors ensures self-aligned beam retracing, thereby significantly simplifying alignment and improving long-term phase stability. The interferometers are configured with a short (S) and a long (L) arm, defining an optical path-length difference of $\Delta L = 43$~cm, corresponding to a temporal delay of $\Delta T = 1.4$~ns (with typical arm lengths of $L = 25$~cm and $S = 3.5$~cm). The short-arm mirrors and one long-arm mirror are mounted on translation stages for coarse path-length matching, while both short-arm mirrors are additionally equipped with piezoelectric actuators, enabling nanometer-scale phase control. The XX and X emission channels, spectrally filtered by the transmission spectrometer, are coupled into the interferometers via single-mode fibers. After recombination at the beamsplitter, emission at both output ports of each interferometer is collected using fiber couplers and routed to a four-channel SNSPD system. Detection events are time-tagged using a time-correlated single-photon counting (TCSPC) module, providing a combined temporal resolution of approximately 100~ps and a system detection efficiency of approximately 85\%. In an initial configuration, only one output port per interferometer (CH1, CH2) was monitored, resulting in a 50\% reduction in usable signal. The setup was subsequently upgraded with additional fiber couplers and routing optics, enabling full four-channel detection (CH1--CH4) and improved coincidence efficiency.

\begin{figure*}
 \centering
 \includegraphics[height=12cm]{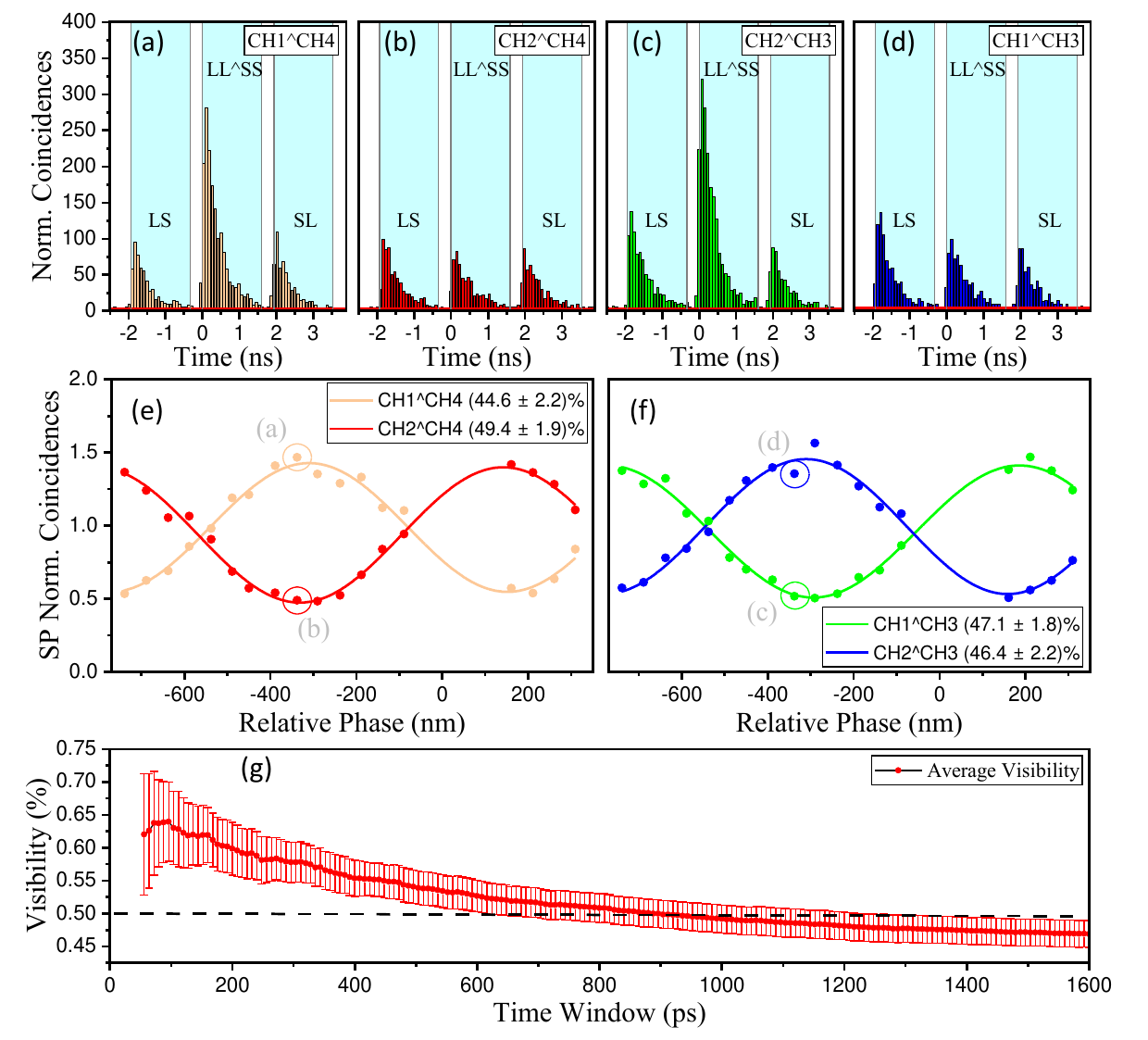}
 \caption{Franson interference measurements of the investigated QD grown on silicon, performed under two-photon resonant excitation at $P = 0.023\,P_{\mathrm{sat}}$. (a–d) Coincidence histograms recorded at interference maxima and minima of the $\mathrm{LL/SS}$ path contributions, shown with a temporal binning of 80 ps. The cyan-shaded region denotes the 1600 ps integration window used to extract the interference signal, while the red line indicates the non-interfering background contribution. (e,f) Phase-dependent $\mathrm{LL/SS}$ coincidence counts, normalized to the side peaks (SPs), exhibiting high-contrast sinusoidal oscillations (solid lines: fits), with extracted visibilities indicated in the legend. (g) Mean two-photon interference visibility as a function of integration window width. The dashed line indicates the 50\% level, which is the threshold for identifying non-classical light emission.}
 \label{Fig:Vis}
\end{figure*}

For energy--time entanglement characterization, coincidence correlations between the interferometer outputs are recorded, yielding the four relevant pairwise combinations CH1\textasciicircum CH3, CH1\textasciicircum CH4, CH2\textasciicircum CH3, and CH2\textasciicircum CH4. Second-order interference is obtained by scanning the total interferometer phase via piezoelectric modulation of the short-arm lengths. The coincidence counts are recorded as a function of phase using an automated data acquisition system.

Due to the low signal level associated with single-photon emission measurements, long-term interferometric stability is required. To achieve this, we employ active feedback stabilization using a reference laser integrated into the same setup, enabling phase stability over several hours with a precision on the order of tens of nanometers. Phase-stabilization is implemented by using a reference laser (CW diode laser) featuring a coherence length of several kilometers which is fiber-coupled and split into both Michelson interferometers via a fiber beamsplitter. The interference visibility of the stabilization signal depends sensitively on the laser polarization, since orthogonally polarized components do not interfere. Therefore, a fiber polarization controller (fiber paddle) was inserted into the reference-laser fiber path to optimize the polarization state. The laser is then out-coupled along a parallel path (green in Fig.~\ref{Fig:Setup}) relative to the signal path, and The first-order interference signal was detected by two silicon single-photon avalanche diodes (SPADs; Excelitas single-photon counting modules) and recorded with a time tagger, which provide the feedback signal for a PID controller. Each interferometer operates with an independent PID loop controlling the piezo actuator in the short arm.

\section{Energy--time entanglement of the biexciton-exciton cascade}

In the following section, we investigate the degree of energy--time entanglement of the X--XX cascade emitted from the QD under study by means of Franson interferometry. To suppress excitation-induced dephasing (EID) and re-excitation processes, both of which are known to significantly limit Franson interference visibility~\cite{Monniello2013,Hohn2023}, the measurements were performed at a low excitation power of \(0.023\,P_{\mathrm{sat}}\). Following a complete phase scan across one full interference period, the coincidence histograms recorded at the maximum and minimum interference conditions are presented in Fig.~\ref{Fig:Vis}(a--d), using a temporal bin width of 80~ps. The post-selection windows are indicated by the blue-shaded regions, with a width of 1600~ps applied to both the central and SPs. At the observed low coincidence rates, the red line denotes the non-interfering background contribution. Importantly, the interfering LL/SS events are well separated from the adjacent SL and LS peaks, enabling unambiguous temporal post-selection of the entangled photon pairs. Note, the data were post-selected in two steps. First, coincidence events corresponding to entangled photon pairs were selected within a temporal window of 0--1600 ps (blue-shaded regions). Second, the non-interfering background contribution was subtracted, as indicated by the red line.

To extract the interference visibility, the coincidence counts within the central peak (central blue-shaded region) were integrated for each relative phase setting over the interval from \(-690\)~nm to \(300\)~nm and subsequently normalized to the corresponding side-peak coincidences (side blue-shaded regions in Fig. \ref{Fig:Vis}(a-d)). The results of this analysis are presented in Figure~\ref{Fig:Vis}(e) and (f) for an integration window of 1600~ps. The solid colored curves represent sinusoidal fits used to determine the visibility for the respective channel combinations. A pronounced variation in the extracted visibilities is observed between different detector-channel combinations, ranging from a maximum of \((49.4 \pm 1.9)\%\) for \(\mathrm{CH2}^{\wedge}\mathrm{CH4}\) to a minimum of \((44.6 \pm 2.2)\%\) for \(\mathrm{CH1}^{\wedge}\mathrm{CH4}\). Under ideal interferometric conditions, such deviations are not expected and therefore indicate residual path imbalances, most likely originating from non-ideal beamsplitter ratios within the interferometer arms. In this context, the highest measured visibility of \((49.4 \pm 1.9)\%\) obtained for \(\mathrm{CH2}^{\wedge}\mathrm{CH4}\) constitutes the least imbalance affected estimate of the intrinsic two-photon interference visibility, since path imbalances can only reduce the experimentally observed visibility during a full phase scan and cannot artificially enhance it. 

Nevertheless, for consistency, we report the average visibility obtained from the two detector-channel combinations using an integration window of 1600~ps. The resulting Franson interference visibility as a function of the chosen temporal integration window is presented in Fig.~\ref{Fig:Vis}(g). We observe a continuous reduction in the extracted interference visibility from \((64 \pm 7)\%\) for an 80~ps integration window to \((47 \pm 2)\%\) for a 1600~ps window. Previous studies have attributed such behavior to an increasing contribution of background events at larger temporal integration windows~\cite{Hohn2023,Xavier2025}. In these reports, background subtraction was shown to render the extracted visibility largely independent of the selected window width. In contrast, the data presented in Figure~\ref{Fig:Vis} already include background subtraction, yet the reduction in visibility persists. This observation therefore indicates the presence of a residual non-interfering contribution within the selected coincidence peak. 

Several mechanisms may account for this behavior. First, systematic uncertainties in the determination of the background level from fits to the long-time bunching contribution become increasingly significant for larger integration windows. While the correlated coincidence signal decays with increasing delay time, any residual offset remains effectively constant and therefore contributes more strongly to the integrated counts at wider windows. Notably, the long-time bunching observed in monolithically integrated InGaAs/GaAs QDs on silicon is more accurately described by a Gaussian-like contribution, in contrast to the exponential behavior commonly reported for InGaAs/GaAs QDs grown on native GaAs substrates \cite{Davano2014}. This deviation suggests the presence of more intricate background dynamics that are difficult to fully disentangle from the interferometric cascade signal. Consequently, a small residual background component may remain after subtraction and progressively reduce the extracted visibility as the integration window is increased.

A further aspect relevant to the Franson interference experiment is the relation between the interferometer delay and the radiative lifetimes of the emitter. The path imbalance has to be sufficiently large to temporally separate the non-interfering SL and LS contributions from the central SS--LL peak and to suppress residual single-photon interference. At the same time, the interfering two-photon amplitudes must retain phase coherence over this delay. The radiative lifetimes of the XX and X transitions determine the temporal extent of the cascade wavepacket and therefore set the natural time scale on which the interferometer delay has to be chosen. For the present emitter, with decay times of $(286 \pm 12)$~ps for XX and $(574 \pm 45)$~ps for X, a delay on the order of several nanoseconds imposes a stringent coherence requirement and may reduce the effective overlap of the interfering SS and LL contributions. A shorter path imbalance could therefore be beneficial, provided that the central and SPs remain clearly distinguishable. However, this cannot be verified from the present data alone and should be regarded as an optimization route for future experiments rather than as a conclusive explanation of the reduced visibility.

The measured Franson interference visibility of up to $(64 \pm 7)\%$ for an 80~ps integration window demonstrates a competitive degree of two-photon coherence in monolithically integrated InGaAs/GaAs QDs on silicon. In comparison to state-of-the-art III/V QD sources, which typically report Franson or time-bin visibilities in the range of $\approx 71\%$ \cite{Hohn2023} to $84\%$ \cite{Gins2021}, depending on excitation scheme, spectral filtering, and interferometric stability, our results place the present platform within the established performance window for non-classical energy--time entangled photon generation, despite the additional complexity introduced by heterogeneous integration on silicon. Importantly, while record-high visibilities in optimized InAs/GaAs systems grown on native substrates often rely on highly engineered photonic structures and ultra-low background conditions, the present device operates under monolithic integration constraints that are inherently more demanding, yet still preserves a clear non-classical interference signature. The observed reduction in visibility with increasing integration window, down to $(47 \pm 2)\%$ at 1600~ps, further distinguishes the present system from idealized literature benchmarks, where background-subtracted visibilities are typically largely invariant with respect to the post-selection window size. This deviation highlights that, in integrated III/V-on-Si platforms, residual non-interfering contributions and interferometer imperfections are more pronounced than in bulk or hybrid fiber-based implementations.

From the perspective of quantum information technology, these results are particularly significant. Energy--time entangled photon pairs generated in a silicon-compatible material system provide a direct pathway toward on-chip quantum communication architectures that are inherently resilient to the FSS of the quantum emitter. Moreover, the monolithic integration of the emitter on silicon enables deterministic coupling to photonic circuits, offering a route toward large-scale quantum photonic networks, multiplexed entangled photon sources, and integrated quantum repeater nodes. In particular, the demonstrated coherence of the XX--X cascade under TPE supports its applicability for heralded entanglement distribution protocols and time-bin encoded quantum key distribution schemes.

Looking ahead, several concrete optimization pathways can be implemented. In particular, improved interferometric balance, reduced residual background emission, and tailoring of the interferometer delay to more closely match the radiative lifetimes of the XX and X states are expected to significantly enhance the achievable visibility and may enable violation of a Bell inequality in future device iterations. Combined with ongoing advances in the epitaxial quality of QDs grown on silicon, these improvements position epitaxial InGaAs/GaAs QDs on silicon as a promising candidate for scalable quantum photonic hardware, bridging the gap between high-performance quantum emitters and industrial silicon technology.

\section*{Conclusions}

Overall, the results obtained highlight the strong potential of epitaxially grown InGaAs/GaAs QDs on silicon as a scalable platform for the generation of energy--time entangled photon pairs and the realization of coherent quantum photonic technologies. By combining monolithic QD integration with coherent excitation of the XX--X cascade, the present platform enables the generation of non-classical light states that are intrinsically robust against FSS and therefore circumvent a major limitation of polarization-entangled photon sources. The observation of Franson interference visibilities of up to $(64 \pm 7)\%$, from epitaxially grown InGaAs/GaAs QDs on silicon demonstrates the preservation of two-photon coherence throughout the biexciton--exciton radiative cascade and confirms the ability of this platform to generate non-classical photonic states under coherent excitation. 

Although the observed visibility remains below the threshold for Bell inequality violation due to residual non-interfering contributions, our analysis indicates that these limitations arise predominantly from technical and experimental factors rather than from fundamental constraints of the developed III/V-on-silicon material platform. Further optimization of the interferometric architecture, background suppression, and matching of the interferometer delay to the emitter dynamics is therefore expected to yield significant improvements in the attainable degree of entanglement. Taken together, our findings establish monolithically integrated InGaAs/GaAs QDs on silicon as a promising and CMOS-compatible route toward scalable quantum photonic circuits capable of generating coherent and entangled photonic states for applications in quantum communication, quantum networking, and distributed quantum information processing.

\section*{Author contributions}

M.~H. designed and built the four-channel Franson interferometer, performed the $\mu$PL experiments, and carried out the data analysis and led evaluation of the results. I.~L. epitaxially grew the upper part of the sample, including the QDs and DBR mirror. P.~L. performed the growth and optimization of the initial GaP/Si template. C.~C.~P. and A.~K.-S. contributed to the interpretation of the results and supported the preparation of the manuscript. K.~V., S.~R., and I.~L. conceived and supervised the project. The manuscript was written by K.~V., S.~R., and I.~L. with contributions from all authors.

\section*{Conflicts of interest}

The authors have no conflicts to disclose.

\section*{Data availability}

The data that support the findings of this study are available from the corresponding author upon reasonable request.

\section*{Acknowledgements}

The research leading to these results received funding from the Federal Ministry of Research, Technology and Space (BMFTR) through projects 16KISQ014 and 16KISQ087K, from the German Research Foundation (DFG) through projects RE2974/23-1 and INST 131/795-1 FUGG, and from the Berlin Senate through Berlin Quantum (BQ). The authors also thank Kathrin Schatke, Praphat Sonka, Heike Oppermann, and Stefan Bock for their technical support. Furthermore, the authors acknowledge Martin von Helversen and Daniel Vajner for their invaluable support and stimulating scientific discussions.


\bibliography{References}

@article{Young_2006,
doi = {10.1088/1367-2630/8/2/029},
url = {https://doi.org/10.1088/1367-2630/8/2/029},
year = {2006},
month = {feb},
publisher = {IOP Publishing},
volume = {8},
number = {2},
pages = {29},
author = {Young, Robert J and Stevenson, R Mark and Atkinson, Paola and Cooper, Ken and Ritchie, David A and Shields, Andrew J},
title = {Improved fidelity of triggered entangled photons from single quantum dots},
journal = {New Journal of Physics}
}

@article{Dousse2008,
  title = {Controlled Light-Matter Coupling for a Single Quantum Dot Embedded in a Pillar Microcavity Using Far-Field Optical Lithography},
  author = {Dousse, A. and Lanco, L. and Suffczy\'{n}ski, J. and Semenova, E. and Miard, A. and Lema{\^i}tre, A. and Sagnes, I. and Roblin, C. and Bloch, J. and Senellart, P.},
  journal = {Physical Review Letters},
  volume = {101},
  issue = {26},
  pages = {267404},
  numpages = {4},
  year = {2008},
  month = {Dec},
  publisher = {American Physical Society},
  doi = {10.1103/PhysRevLett.101.267404},
  url = {https://link.aps.org/doi/10.1103/PhysRevLett.101.267404}
}

@article{PELUCCHI2004476,
title = {Site-controlled quantum dots grown in inverted pyramids for photonic crystal applications},
journal = {Physica E: Low-dimensional Systems and Nanostructures},
volume = {23},
number = {3},
pages = {476-481},
year = {2004},
note = {Proceedings of the Fifth International Workshop on Epitaxial Semiconductors on Patterned Substrates and Novel Index Surfaces (ESPS-NIS)},
issn = {1386-9477},
doi = {https://doi.org/10.1016/j.physe.2004.02.010},
url = {https://www.sciencedirect.com/science/article/pii/S1386947704001699},
author = {E Pelucchi and S Watanabe and K Leifer and B Dwir and E Kapon}
}

@article{Zopf2019,
  title = {Entanglement Swapping with Semiconductor-Generated Photons Violates Bell's Inequality},
  author = {Zopf, Michael and Keil, Robert and Chen, Yan and Yang, Jingzhong and Chen, Disheng and Ding, Fei and Schmidt, Oliver G.},
  journal = {Physical Review Letters},
  volume = {123},
  issue = {16},
  pages = {160502},
  numpages = {7},
  year = {2019},
  month = {Oct},
  publisher = {American Physical Society},
  doi = {10.1103/PhysRevLett.123.160502},
  url = {https://link.aps.org/doi/10.1103/PhysRevLett.123.160502}
}

@article{Waks2002,
  title = {Quantum cryptography with a photon turnstile},
  volume = {420},
  ISSN = {1476-4687},
  url = {http://dx.doi.org/10.1038/420762a},
  DOI = {10.1038/420762a},
  number = {6917},
  journal = {Nature},
  publisher = {Springer Science and Business Media LLC},
  author = {Waks,  Edo and Inoue,  Kyo and Santori,  Charles and Fattal,  David and Vuckovic,  Jelena and Solomon,  Glenn S. and Yamamoto,  Yoshihisa},
  year = {2002},
  month = Dec,
  pages = {762–762}
}

@article{Heindel2012,
  title = {Quantum key distribution using quantum dot single-photon emitting diodes in the red and near infrared spectral range},
  volume = {14},
  ISSN = {1367-2630},
  url = {http://dx.doi.org/10.1088/1367-2630/14/8/083001},
  DOI = {10.1088/1367-2630/14/8/083001},
  number = {8},
  journal = {New Journal of Physics},
  publisher = {IOP Publishing},
  author = {Heindel,  Tobias and Kessler,  Christian A and Rau,  Markus and Schneider,  Christian and F\"{u}rst,  Martin and Hargart,  Fabian and Schulz,  Wolfgang-Michael and Eichfelder,  Marcus and Roßbach,  Robert and Nauerth,  Sebastian and Lermer,  Matthias and Weier,  Henning and Jetter,  Michael and Kamp,  Martin and Reitzenstein,  Stephan and H\"{o}fling,  Sven and Michler,  Peter and Weinfurter,  Harald and Forchel,  Alfred},
  year = {2012},
  month = Aug,
  pages = {083001}
}

@incollection{Reitzenstein_CSSCST2024,
  author    = {Reitzenstein, Stephan},
  title     = {Generation of indistinguishable photons with semiconductor quantum dots},
  booktitle = {Comprehensive Semiconductor Science and Technology},
  editor    = {Fornari, Roberto and Grundmann, Marius and Darakchieva, Vanya and Fay, Patrick},
  volume    = {2},
  series    = {Comprehensive Semiconductor Science and Technology},
  publisher = {Elsevier},
  year      = {2024},
  address   = {Amsterdam},
  doi       = {10.1016/B978-0-323-96027-4.<chapter-DOI>},
}

@article{vonHelversen2022,
author = {von Helversen, Martin and Haisler, Alexey Vladimirovich and Daurtsev, Marat Petrovich and Dmitriev, Dmitry Vladimirovich and Toropov, Alexander Ivanovich and Rodt, Sven and Haisler, Vladimir Anatolievich and Derebezov, Ilya Alexandrovich and Reitzenstein, Stephan},
title = {Triggered Single-Photon Emission of Resonantly Excited Quantum Dots Grown on (111)B GaAs Substrate},
journal = {physica status solidi (RRL) – Rapid Research Letters},
volume = {16},
number = {8},
pages = {2200133},
doi = {https://doi.org/10.1002/pssr.202200133},
url = {https://onlinelibrary.wiley.com/doi/abs/10.1002/pssr.202200133},
eprint = {https://onlinelibrary.wiley.com/doi/pdf/10.1002/pssr.202200133},
year = {2022}
}

@article{Yang:24,
author = {Jiawei Yang and Zhixuan Rao and Changkun Song and Mujie Rao and Ziyang Zheng and Luyu Liu and Xuebin Peng and Ying Yu and Siyuan Yu},
journal = {Photonics Research},
number = {10},
pages = {2130--2138},
publisher = {Optica Publishing Group},
title = {Filter-free high-performance single-photon emission from a quantum dot in a Fabry-Perot microcavity},
volume = {12},
month = {Oct},
year = {2024},
url = {https://opg.optica.org/prj/abstract.cfm?URI=prj-12-10-2130},
doi = {10.1364/PRJ.523970},
}

@article{Limame2024,
  title = {High-quality single InGaAs/GaAs quantum dot growth on a silicon substrate for quantum photonic applications},
  volume = {2},
  ISSN = {2837-6714},
  url = {http://dx.doi.org/10.1364/OPTICAQ.510829},
  DOI = {10.1364/opticaq.510829},
  number = {2},
  journal = {Optica Quantum},
  publisher = {Optica Publishing Group},
  author = {Limame,  Imad and Ludewig,  Peter and Shih,  Ching-Wen and Hohn,  Marcel and Palekar,  Chirag C. and Stolz,  Wolfgang and Reitzenstein,  Stephan},
  year = {2024},
  month = Apr,
  pages = {117}
}

@article{Heindel2023,
  title = {Quantum dots for photonic quantum information technology},
  volume = {15},
  ISSN = {1943-8206},
  url = {http://dx.doi.org/10.1364/AOP.490091},
  DOI = {10.1364/aop.490091},
  number = {3},
  journal = {Advances in Optics and Photonics},
  publisher = {Optica Publishing Group},
  author = {Heindel,  Tobias and Kim,  Je-Hyung and Gregersen,  Niels and Rastelli,  Armando and Reitzenstein,  Stephan},
  year = {2023},
  month = Aug,
  pages = {613}
}

@article{Senellart2017,
  title = {High-performance semiconductor quantum-dot single-photon sources},
  volume = {12},
  ISSN = {1748-3395},
  url = {http://dx.doi.org/10.1038/nnano.2017.218},
  DOI = {10.1038/nnano.2017.218},
  number = {11},
  journal = {Nature Nanotechnology},
  publisher = {Springer Science and Business Media LLC},
  author = {Senellart,  Pascale and Solomon,  Glenn and White,  Andrew},
  year = {2017},
  month = Nov,
  pages = {1026–1039}
}

@article{Esmann2024,
  title = {Solid‐State Single‐Photon Sources: Recent Advances for Novel Quantum Materials},
  volume = {34},
  ISSN = {1616-3028},
  url = {http://dx.doi.org/10.1002/adfm.202315936},
  DOI = {10.1002/adfm.202315936},
  number = {30},
  journal = {Advanced Functional Materials},
  publisher = {Wiley},
  author = {Esmann,  Martin and Wein,  Stephen C. and Antón‐Solanas,  Carlos},
  year = {2024},
  month = Apr,
  pages = {2315936}
}

@article{Chen2024,
  title = {Wavelength-tunable high-fidelity entangled photon sources enabled by dual Stark effects},
  volume = {15},
  ISSN = {2041-1723},
  url = {http://dx.doi.org/10.1038/s41467-024-50062-0},
  DOI = {10.1038/s41467-024-50062-0},
  number = {1},
  journal = {Nature Communications},
  publisher = {Springer Science and Business Media LLC},
  author = {Chen,  Chen and Yan,  Jun-Yong and Babin,  Hans-Georg and Wang,  Jiefei and Xu,  Xingqi and Lin,  Xing and Yu,  Qianqian and Fang,  Wei and Liu,  Run-Ze and Huo,  Yong-Heng and Cai,  Han and Sha,  Wei E. I. and Zhang,  Jiaxiang and Heyn,  Christian and Wieck,  Andreas D. and Ludwig,  Arne and Wang,  Da-Wei and Jin,  Chao-Yuan and Liu,  Feng},
  year = {2024},
  month = July,
  pages = {5792}
}

@article{Huber2017,
  title = {Highly indistinguishable and strongly entangled photons from symmetric GaAs quantum dots},
  volume = {8},
  ISSN = {2041-1723},
  url = {http://dx.doi.org/10.1038/ncomms15506},
  DOI = {10.1038/ncomms15506},
  number = {1},
  journal = {Nature Communications},
  publisher = {Springer Science and Business Media LLC},
  author = {Huber,  Daniel and Reindl,  Marcus and Huo,  Yongheng and Huang,  Huiying and Wildmann,  Johannes S. and Schmidt,  Oliver G. and Rastelli,  Armando and Trotta,  Rinaldo},
  year = {2017},
  month = May,
  pages = {15506} 
}

@article{Hauser2026,
  title = {Deterministic and highly indistinguishable single photons in the telecom C-band},
  volume = {17},
  ISSN = {2041-1723},
  url = {http://dx.doi.org/10.1038/s41467-026-68336-0},
  DOI = {10.1038/s41467-026-68336-0},
  number = {1},
  journal = {Nature Communications},
  publisher = {Springer Science and Business Media LLC},
  author = {Hauser,  Nico and Bayerbach,  Matthias and Kaupp,  Jochen and Reum,  Yorick and Peniakov,  Giora and Michl,  Johannes and Kamp,  Martin and Huber-Loyola,  Tobias and Pfenning,  Andreas T. and H\"{o}fling,  Sven and Barz,  Stefanie},
  year = {2026},
  month = Jan,
  pages = {537}
}

@article{Hohn2023,
  title = {Energy-time entanglement from a resonantly driven quantum-dot three-level system},
  volume = {5},
  ISSN = {2643-1564},
  url = {http://dx.doi.org/10.1103/PhysRevResearch.5.L022060},
  DOI = {10.1103/physrevresearch.5.l022060},
  number = {2},
  journal = {Physical Review Research},
  publisher = {American Physical Society (APS)},
  author = {Hohn,  M. and Barkemeyer,  K. and von Helversen,  M. and Bremer,  L. and Gschrey,  M. and Schulze,  J.-H. and Strittmatter,  A. and Carmele,  A. and Rodt,  S. and Bounouar,  S. and Reitzenstein,  S.},
  year = {2023},
  month = June,
  pages = {L022060}
}

@article{Ding2016,
  title = {On-Demand Single Photons with High Extraction Efficiency and Near-Unity Indistinguishability from a Resonantly Driven Quantum Dot in a Micropillar},
  volume = {116},
  ISSN = {1079-7114},
  url = {http://dx.doi.org/10.1103/PhysRevLett.116.020401},
  DOI = {10.1103/physrevlett.116.020401},
  number = {2},
  journal = {Physical Review Letters},
  publisher = {American Physical Society (APS)},
  author = {Ding,  Xing and He,  Yu and Duan,  Z.-C. and Gregersen,  Niels and Chen,  M.-C. and Unsleber,  S. and Maier,  S. and Schneider,  Christian and Kamp,  Martin and H\"{o}fling,  Sven and Lu,  Chao-Yang and Pan,  Jian-Wei},
  year = {2016},
  month = Jan,
  pages = {020401}
}

@article{Sartison2018,
  title = {Deterministic integration and optical characterization of telecom O-band quantum dots embedded into wet-chemically etched Gaussian-shaped microlenses},
  volume = {113},
  ISSN = {1077-3118},
  url = {http://dx.doi.org/10.1063/1.5038271},
  DOI = {10.1063/1.5038271},
  number = {3},
  journal = {Applied Physics Letters},
  publisher = {AIP Publishing},
  author = {Sartison,  Marc and Engel,  Lena and Kolatschek,  Sascha and Olbrich,  Fabian and Nawrath,  Cornelius and Hepp,  Stefan and Jetter,  Michael and Michler,  Peter and Portalupi,  Simone Luca},
  year = {2018},
  month = July,
  pages = {032103}
}

@article{BassoBasset2019,
  title = {Entanglement Swapping with Photons Generated on Demand by a Quantum Dot},
  volume = {123},
  ISSN = {1079-7114},
  url = {http://dx.doi.org/10.1103/PhysRevLett.123.160501},
  DOI = {10.1103/physrevlett.123.160501},
  number = {16},
  journal = {Physical Review Letters},
  publisher = {American Physical Society (APS)},
  author = {Basso Basset,  F. and Rota,  M. B. and Schimpf,  C. and Tedeschi,  D. and Zeuner,  K. D. and Covre da Silva,  S. F. and Reindl,  M. and Zwiller,  V. and J\"{o}ns,  K. D. and Rastelli,  A. and Trotta,  R.},
  year = {2019},
  month = Oct,
  pages = {160501}
}

@article{Gaur2026,
  title = {Scalable quantum photonic platform based on site-controlled quantum dots coupled to circular Bragg grating resonators},
  volume = {15},
  ISSN = {2047-7538},
  url = {http://dx.doi.org/10.1038/s41377-026-02343-0},
  DOI = {10.1038/s41377-026-02343-0},
  number = {1},
  journal = {Light: Science \& Applications},
  publisher = {Springer Science and Business Media LLC},
  author = {Gaur,  Kartik and Barua,  Avijit and Tripathi,  Sarthak and Roche,  Léo J. and Wilksen,  Steffen and Steinhoff,  Alexander and Baraz,  Sam and Nitin,  Neha and Palekar,  Chirag C. and Koulas-Simos,  Aris and Limame,  Imad and Mudi,  Priyabrata and Rodt,  Sven and Gies,  Christopher and Reitzenstein,  Stephan},
  year = {2026},
  month = June,
  pages = {02343}
}

@article{Schneider2012,
  title = {In(Ga)As/GaAs site‐controlled quantum dots with tailored morphology and high optical quality},
  volume = {209},
  ISSN = {1862-6319},
  url = {http://dx.doi.org/10.1002/pssa.201228373},
  DOI = {10.1002/pssa.201228373},
  number = {12},
  journal = {physica status solidi (a)},
  publisher = {Wiley},
  author = {Schneider,  Christian and Huggenberger,  Alexander and Gschrey,  Manuel and Gold,  Peter and Rodt,  Sven and Forchel,  Alfred and Reitzenstein,  Stephan and H\"{o}fling,  Sven and Kamp,  Martin},
  year = {2012},
  month = Sept,
  pages = {2379–2386}
}

@article{Akopian2006,
  title = {Entangled Photon Pairs from Semiconductor Quantum Dots},
  volume = {96},
  ISSN = {1079-7114},
  url = {http://dx.doi.org/10.1103/PhysRevLett.96.130501},
  DOI = {10.1103/physrevlett.96.130501},
  number = {13},
  journal = {Physical Review Letters},
  publisher = {American Physical Society (APS)},
  author = {Akopian,  N. and Lindner,  N. H. and Poem,  E. and Berlatzky,  Y. and Avron,  J. and Gershoni,  D. and Gerardot,  B. D. and Petroff,  P. M.},
  year = {2006},
  month = Apr,
  pages = {130501}
}

@article{Xavier2025,
  title = {Energy-time and time-bin entanglement: past,  present and future},
  volume = {11},
  ISSN = {2056-6387},
  url = {http://dx.doi.org/10.1038/s41534-025-01072-3},
  DOI = {10.1038/s41534-025-01072-3},
  number = {1},
  journal = {npj Quantum Information},
  publisher = {Springer Science and Business Media LLC},
  author = {Xavier, Guilherme B. and Larsson, Jan-{\AA}ke and Villoresi, Paolo and Vallone, Giuseppe and Cabello, Ad{\'a}n},
  year = {2025},
  month = July,
  pages = {01072}
}

@article{Yerino2014,
  title = {Strain-driven growth of GaAs(111) quantum dots with low fine structure splitting},
  volume = {105},
  ISSN = {1077-3118},
  url = {http://dx.doi.org/10.1063/1.4904944},
  DOI = {10.1063/1.4904944},
  number = {25},
  journal = {Applied Physics Letters},
  publisher = {AIP Publishing},
  author = {Yerino,  Christopher D. and Simmonds,  Paul J. and Liang,  Baolai and Jung,  Daehwan and Schneider,  Christian and Unsleber,  Sebastian and Vo,  Minh and Huffaker,  Diana L. and H\"{o}fling,  Sven and Kamp,  Martin and Lee,  Minjoo Larry},
  year = {2014},
  month = Dec,
  pages = {251901}
}

@article{Huo2013,
  title = {Ultra-small excitonic fine structure splitting in highly symmetric quantum dots on GaAs (001) substrate},
  volume = {102},
  ISSN = {1077-3118},
  url = {http://dx.doi.org/10.1063/1.4802088},
  DOI = {10.1063/1.4802088},
  number = {15},
  journal = {Applied Physics Letters},
  publisher = {AIP Publishing},
  author = {Huo,  Y. H. and Rastelli,  A. and Schmidt,  O. G.},
  year = {2013},
  month = Apr,
  pages = {152105}
}

@article{Sun2017,
  title = {Measurement of the inhomogeneous broadening of a bi-exciton state in a quantum dot using Franson-type nonlocal interference},
  volume = {25},
  ISSN = {1094-4087},
  url = {http://dx.doi.org/10.1364/OE.25.001778},
  DOI = {10.1364/oe.25.001778},
  number = {3},
  journal = {Optics Express},
  publisher = {Optica Publishing Group},
  author = {Sun,  Yong-Nan and Zou,  Yang and Chen,  Geng and Tang,  Jian-Shun and Ni,  Hai-Qiao and Li,  Mi-Feng and Zha,  Guo-Wei and Niu,  Zhi-Chuan and Han,  Yong-Jian and Li,  Chuan-Feng and Guo,  Guang-Can},
  year = {2017},
  month = Jan,
  pages = {1778}
}

@article{Gins2021,
  title = {Time-bin entangled photon pairs from quantum dots embedded in a self-aligned cavity},
  volume = {29},
  ISSN = {1094-4087},
  url = {http://dx.doi.org/10.1364/OE.411021},
  DOI = {10.1364/oe.411021},
  number = {3},
  journal = {Optics Express},
  publisher = {Optica Publishing Group},
  author = {Ginés,  Laia and Pepe,  Carlo and Gonzales,  Junior and Gregersen,  Niels and H\"{o}fling,  Sven and Schneider,  Christian and Predojević,  Ana},
  year = {2021},
  month = Jan,
  pages = {4174}
}

@article{Jayakumar2014,
  title = {Time-bin entangled photons from a quantum dot},
  volume = {5},
  ISSN = {2041-1723},
  url = {http://dx.doi.org/10.1038/ncomms5251},
  DOI = {10.1038/ncomms5251},
  number = {1},
  journal = {Nature Communications},
  publisher = {Springer Science and Business Media LLC},
  author = {Jayakumar,  Harishankar and Predojević,  Ana and Kauten,  Thomas and Huber,  Tobias and Solomon,  Glenn S. and Weihs,  Gregor},
  year = {2014},
  month = June,
  pages = {4251}
}

@article{Lee2019,
  title = {A quantum dot as a source of time-bin entangled multi-photon states},
  volume = {4},
  ISSN = {2058-9565},
  url = {http://dx.doi.org/10.1088/2058-9565/ab0a9b},
  DOI = {10.1088/2058-9565/ab0a9b},
  number = {2},
  journal = {Quantum Science and Technology},
  publisher = {IOP Publishing},
  author = {Lee,  J P and Villa,  B and Bennett,  A J and Stevenson,  R M and Ellis,  D J P and Farrer,  I and Ritchie,  D A and Shields,  A J},
  year = {2019},
  month = Mar,
  pages = {025011}
}

@article{Volz2011,
  title = {GaP-nucleation on exact Si (001) substrates for III/V device integration},
  volume = {315},
  ISSN = {0022-0248},
  url = {http://dx.doi.org/10.1016/j.jcrysgro.2010.10.036},
  DOI = {10.1016/j.jcrysgro.2010.10.036},
  number = {1},
  journal = {Journal of Crystal Growth},
  publisher = {Elsevier BV},
  author = {Volz,  Kerstin and Beyer,  Andreas and Witte,  Wiebke and Ohlmann,  Jens and Németh,  Igor and Kunert,  Bernardette and Stolz,  Wolfgang},
  year = {2011},
  month = Jan,
  pages = {37–47}
}

@article{Beyer2013,
  title = {Atomic structure of (110) anti-phase boundaries in GaP on Si(001)},
  volume = {103},
  ISSN = {1077-3118},
  url = {http://dx.doi.org/10.1063/1.4815985},
  DOI = {10.1063/1.4815985},
  number = {3},
  journal = {Applied Physics Letters},
  publisher = {AIP Publishing},
  author = {Beyer,  A. and Haas,  B. and Gries,  K. I. and Werner,  K. and Luysberg,  M. and Stolz,  W. and Volz,  K.},
  year = {2013},
  month = July,
  pages = {032107}
}

@article{SnchezMuoz2015,
  title = {Enhanced two-photon emission from a dressed biexciton},
  volume = {17},
  ISSN = {1367-2630},
  url = {http://dx.doi.org/10.1088/1367-2630/17/12/123021},
  DOI = {10.1088/1367-2630/17/12/123021},
  number = {12},
  journal = {New Journal of Physics},
  publisher = {IOP Publishing},
  author = {Sánchez Muñoz,  Carlos and Laussy,  Fabrice P and Tejedor,  Carlos and Valle,  Elena del},
  year = {2015},
  month = Dec,
  pages = {123021}
}

@article{Jundt2008,
  title = {Observation of Dressed Excitonic States in a Single Quantum Dot},
  volume = {100},
  ISSN = {1079-7114},
  url = {http://dx.doi.org/10.1103/PhysRevLett.100.177401},
  DOI = {10.1103/physrevlett.100.177401},
  number = {17},
  journal = {Physical Review Letters},
  publisher = {American Physical Society (APS)},
  author = {Jundt,  Gregor and Robledo,  Lucio and H\"{o}gele,  Alexander and F\"{a}lt,  Stefan and Imamoğlu,  Atac},
  year = {2008},
  pages = {177401},
  month = Apr 
}

@article{Monniello2013,
  title = {Excitation-Induced Dephasing in a Resonantly Driven InAs/GaAs Quantum Dot},
  volume = {111},
  ISSN = {1079-7114},
  url = {http://dx.doi.org/10.1103/PhysRevLett.111.026403},
  DOI = {10.1103/physrevlett.111.026403},
  number = {2},
  journal = {Physical Review Letters},
  publisher = {American Physical Society (APS)},
  author = {Monniello,  Léonard and Tonin,  Catherine and Hostein,  Richard and Lemaitre,  Aristide and Martinez,  Anthony and Voliotis,  Valia and Grousson,  Roger},
  year = {2013},
  pages = {026403},
  month = July 
}

@misc{Tripathi2026,
  doi = {10.48550/ARXIV.2604.23422},
  url = {https://arxiv.org/abs/2604.23422},
  author = {Tripathi,  Sarthak and Gaur,  Kartik and Mudi,  Priyabrata and Ludewig,  Peter and Kosarev,  Alexander and Volz,  Kerstin and Limame,  Imad and Reitzenstein,  Stephan},
  title = {Stark-tunable O-band single-photon sources based on deterministically fabricated quantum dot--circular Bragg gratings on silicon},
  publisher = {arXiv},
  year = {2026},
  copyright = {Creative Commons Attribution 4.0 International}
}

@article{Schweickert2018,
  title = {On-demand generation of background-free single photons from a solid-state source},
  volume = {112},
  ISSN = {1077-3118},
  url = {http://dx.doi.org/10.1063/1.5020038},
  DOI = {10.1063/1.5020038},
  number = {9},
  journal = {Applied Physics Letters},
  publisher = {AIP Publishing},
  author = {Schweickert,  Lucas and J\"{o}ns,  Klaus D. and Zeuner,  Katharina D. and Covre da Silva,  Saimon Filipe and Huang,  Huiying and Lettner,  Thomas and Reindl,  Marcus and Zichi,  Julien and Trotta,  Rinaldo and Rastelli,  Armando and Zwiller,  Val},
  year = {2018},
  pages = {093106},
  month = Feb 
}

@article{Loredo2017,
  title = {Boson Sampling with Single-Photon Fock States from a Bright Solid-State Source},
  volume = {118},
  ISSN = {1079-7114},
  url = {http://dx.doi.org/10.1103/PhysRevLett.118.130503},
  DOI = {10.1103/physrevlett.118.130503},
  number = {13},
  journal = {Physical Review Letters},
  publisher = {American Physical Society (APS)},
  author = {Loredo,  J.C. and Broome,  M.A. and Hilaire,  P. and Gazzano,  O. and Sagnes,  I. and Lemaitre,  A. and Almeida,  M.P. and Senellart,  P. and White,  A.G.},
  year = {2017},
  pages = {130503},
  month = Mar 
}

@article{Wang2018,
  title = {Toward Scalable Boson Sampling with Photon Loss},
  volume = {120},
  ISSN = {1079-7114},
  url = {http://dx.doi.org/10.1103/PhysRevLett.120.230502},
  DOI = {10.1103/physrevlett.120.230502},
  number = {23},
  journal = {Physical Review Letters},
  publisher = {American Physical Society (APS)},
  author = {Wang,  Hui and Li,  Wei and Jiang,  Xiao and He,  Y.-M. and Li,  Y.-H. and Ding,  X. and Chen,  M.-C. and Qin,  J. and Peng,  C.-Z. and Schneider,  C. and Kamp,  M. and Zhang,  W.-J. and Li,  H. and You,  L.-X. and Wang,  Z. and Dowling,  J. P. and H\"{o}fling,  S. and Lu,  Chao-Yang and Pan,  Jian-Wei},
  year = {2018},
  pages = {230502},
  month = June 
}

@article{Strittmatter2012,
  title = {Site‐controlled quantum dot growth on buried oxide stressor layers},
  volume = {209},
  ISSN = {1862-6319},
  url = {http://dx.doi.org/10.1002/pssa.201228407},
  DOI = {10.1002/pssa.201228407},
  number = {12},
  journal = {physica status solidi (a)},
  publisher = {Wiley},
  author = {Strittmatter,  André and Holzbecher,  André and Schliwa,  Andrei and Schulze,  Jan‐Hindrik and Quandt,  David and Germann,  Tim David and Dreismann,  Alexander and Hitzemann,  Ole and Stock,  Erik and Ostapenko,  Irina A. and Rodt,  Sven and Unrau,  Waldemar and Pohl,  Udo W. and Hoffmann,  Axel and Bimberg,  Dieter and Haisler,  Vladimir},
  year = {2012},
  month = Nov,
  pages = {2411–2420}
}

@article{He2017,
  title = {Deterministic implementation of a bright,  on-demand single-photon source with near-unity indistinguishability via quantum dot imaging},
  volume = {4},
  ISSN = {2334-2536},
  url = {http://dx.doi.org/10.1364/OPTICA.4.000802},
  DOI = {10.1364/optica.4.000802},
  number = {7},
  journal = {Optica},
  publisher = {Optica Publishing Group},
  author = {He,  Yu-Ming and Liu,  Jin and Maier,  Sebastian and Emmerling,  Monika and Gerhardt,  Stefan and Davan\c{c}o,  Marcelo and Srinivasan,  Kartik and Schneider,  Christian and H\"{o}fling,  Sven},
  year = {2017},
  month = July,
  pages = {802}
}

@article{Gschrey2013,
  title = {<i>In situ</i> electron-beam lithography of deterministic single-quantum-dot mesa-structures using low-temperature cathodoluminescence spectroscopy},
  volume = {102},
  ISSN = {1077-3118},
  url = {http://dx.doi.org/10.1063/1.4812343},
  DOI = {10.1063/1.4812343},
  number = {25},
  journal = {Applied Physics Letters},
  publisher = {AIP Publishing},
  author = {Gschrey,  M. and Gericke,  F. and Sch\"{u}ssler,  A. and Schmidt,  R. and Schulze,  J.-H. and Heindel,  T. and Rodt,  S. and Strittmatter,  A. and Reitzenstein,  S.},
  year = {2013},
  month = June,
  pages = {251113}
}

@article{Li2023,
  title = {Scalable Deterministic Integration of Two Quantum Dots into an On-Chip Quantum Circuit},
  volume = {10},
  ISSN = {2330-4022},
  url = {http://dx.doi.org/10.1021/acsphotonics.3c00547},
  DOI = {10.1021/acsphotonics.3c00547},
  number = {8},
  journal = {ACS Photonics},
  publisher = {American Chemical Society (ACS)},
  author = {Li,  Shulun and Yang,  Yuhui and Schall,  Johannes and von Helversen,  Martin and Palekar,  Chirag and Liu,  Hanqing and Roche,  Léo and Rodt,  Sven and Ni,  Haiqiao and Zhang,  Yu and Niu,  Zhichuan and Reitzenstein,  Stephan},
  year = {2023},
  month = July,
  pages = {2846–2853}
}

@article{Benson2000,
  title = {Regulated and Entangled Photons from a Single Quantum Dot},
  volume = {84},
  ISSN = {1079-7114},
  url = {http://dx.doi.org/10.1103/PhysRevLett.84.2513},
  DOI = {10.1103/physrevlett.84.2513},
  number = {11},
  journal = {Physical Review Letters},
  publisher = {American Physical Society (APS)},
  author = {Benson,  Oliver and Santori,  Charles and Pelton,  Matthew and Yamamoto,  Yoshihisa},
  year = {2000},
  month = Mar,
  pages = {2513–2516}
}

@article{Koscica2025,
  title = {Quantum Dot Photodetector and Laser Monolithically Integrated on Silicon Photonics},
  volume = {12},
  ISSN = {2330-4022},
  url = {http://dx.doi.org/10.1021/acsphotonics.5c01299},
  DOI = {10.1021/acsphotonics.5c01299},
  number = {9},
  journal = {ACS Photonics},
  publisher = {American Chemical Society (ACS)},
  author = {Koscica,  Rosalyn and Skipper,  Alec M. and Shi,  Bei and Leake,  Gerald and Zylstra,  Michael and Herman,  Joshua and Liu,  Yuan and Zhang,  Chongxin and Harame,  David and Klamkin,  Jonathan and Bowers,  John},
  year = {2025},
  month = Aug,
  pages = {5173–5178}
}

@article{Bell1964,
  title = {On the Einstein Podolsky Rosen paradox},
  volume = {1},
  ISSN = {0554-128X},
  url = {http://dx.doi.org/10.1103/PhysicsPhysiqueFizika.1.195},
  DOI = {10.1103/physicsphysiquefizika.1.195},
  number = {3},
  journal = {Physics Physique Fizika},
  publisher = {American Physical Society (APS)},
  author = {Bell,  J. S.},
  year = {1964},
  month = Nov,
  pages = {195–200}
}

@article{Davano2014,
  title = {Multiple time scale blinking in InAs quantum dot single-photon sources},
  volume = {89},
  ISSN = {1550-235X},
  url = {http://dx.doi.org/10.1103/PhysRevB.89.161303},
  DOI = {10.1103/physrevb.89.161303},
  number = {16},
  journal = {Physical Review B},
  publisher = {American Physical Society (APS)},
  author = {Davan\c{c}o,  Marcelo and Hellberg,  C. Stephen and Ates,  Serkan and Badolato,  Antonio and Srinivasan,  Kartik},
  year = {2014},
  month = Apr,
  pages = {161303(R)}

}

@misc{ChatGPT2026,
  author       = {{OpenAI}},
  title        = {{ChatGPT (GPT-5.5)}},
  year         = {2026},
  howpublished = {\url{https://chatgpt.com}},
  note         = {Large language model, accessed July 2026}
}

@article{Huber2016,
  title = {Coherence and degree of time-bin entanglement from quantum dots},
  author = {Huber, Tobias and Ostermann, Laurin and Prilm\"uller, Maximilian and Solomon, Glenn S. and Ritsch, Helmut and Weihs, Gregor and Predojevi\ifmmode \acute{c}\else \'{c}\fi{}, Ana},
  journal = {Phys. Rev. B},
  volume = {93},
  issue = {20},
  pages = {201301(R)},
  numpages = {5},
  year = {2016},
  month = {May},
  publisher = {American Physical Society},
  doi = {10.1103/PhysRevB.93.201301},
  url = {https://link.aps.org/doi/10.1103/PhysRevB.93.201301}
}

@article{Reithmaier2004,
  title = {Strong coupling in a single quantum dot–semiconductor microcavity system},
  volume = {432},
  ISSN = {1476-4687},
  url = {http://dx.doi.org/10.1038/nature02969},
  DOI = {10.1038/nature02969},
  number = {7014},
  journal = {Nature},
  publisher = {Springer Science and Business Media LLC},
  author = {Reithmaier,  J. P. and Sęk,  G. and L\"{o}ffler,  A. and Hofmann,  C. and Kuhn,  S. and Reitzenstein,  S. and Keldysh,  L. V. and Kulakovskii,  V. D. and Reinecke,  T. L. and Forchel,  A.},
  year = {2004},
  month = Nov,
  pages = {197–200}
}
\bibliographystyle{aipnum4-1}
\end{document}